\documentclass[fleqn,12pt,twoside]{article}
\usepackage{amsmath}
\usepackage{espcrc1}

\usepackage{graphicx} 
\usepackage[figuresright]{rotating} 
   
\newcommand{\He}{{}^3\mathrm{He}}
\newcommand{\Hh}{{}^3\mathrm{H}}
\newcommand{\etal}{{\em et al.}}

\title{Four-nucleon scattering: a new numerical approach}

\author{A. Deltuva\address[CFNUL]{Centro de F\'\i sica Nuclear, University of Lisbon,\\
    Av. Prof. Gama Pinto 2, 1649-003 Lisbon, Portugal}\thanks
  {Supported by the FCT grant SFRH/BPD/14801/2003}
  and        A.C. Fonseca\addressmark[CFNUL]}
       
\begin{document}

\maketitle

\begin{abstract}

The AGS equations are solved for $n\Hh$ and $p\He$ scattering
including the Coulomb interaction. Comparison with previous work confirms the
accuracy of the calculation and helps clarify a number of issues related to the
$n\Hh$ total cross section at the peak of the resonance region, as well as an
$A_y$ deficiency in $p\He$. Calculations are fully converged in terms of $NN$
partial waves and involve no uncontrolled approximations.

\end{abstract}

\section{INTRODUCTION} 

In spite of the progress that has taken place over the years on the solution
of the three-nucleon scattering problem in terms of ab initio calculations,
there are still some open problems that defy our understanding. They are the
$A_y$ puzzle in elastic $Nd$ scattering and the space star anomaly in $Nd$
breakup, just to name a few. These problems persist even when the Coulomb
interaction is added \cite{Kiev01,Delt05,Delt05a} and seem to be insensitive to the
choice of realistic $2N + 3N$ force model. As it has already been highlighted in the past
\cite{Fon99,Viv01,Lazaus05}, the four-nucleon scattering problem reveals
further discrepancies between theory and experiment that need clarification in
terms of improved calculations using modern force models and
efficient numerical algorithms that allow for a numerically converged solution
of the Alt, Grassberger and Sandhas (AGS) equation \cite{Grass67/72} for the transition
operators.
 
In four-nucleon scattering the Coulomb interaction is paramount to treat
$p\He$, to separate the $n\He$ threshold from $p\Hh$ and at
the same time avoid a second excited state of the alpha particle a few keV bellow the
lowest scattering threshold \cite{Fon02}. Given the success recently achieved
in including the Coulomb interaction in $pd$ elastic scattering and
breakup \cite{Delt05,Delt05a}, we generalize the proposed method to the four-nucleon
system and study $p\He$ below three-body breakup threshold. 

The work we present here constitutes the first effort to design and construct a new
scattering code that may bring the treatment of the four-nucleon system to the same
degree of accuracy and sophistication as we already have in the three-nucleon system. The
symmetrized AGS equations for four identical nucleons are given by
\begin{subequations} \label{eq:U11e21}
\begin{align} \label{eq:U11}
{\cal{U}}^{11}  = {}& -(G_0 \; t \; G_0)^{-1} \; P_{34} -
P_{34} U\; G_0 \; t \;G_0 \ {\cal{U}}^{11} +\tilde{U} \;  G_0 \; t \;G_0 \
{\cal{U}}^{21} ,
\\
\label{eq:U21}
{\cal{U}}^{21}  = {}& (G_0 \; t \; G_0)^{-1} \; (1 - P_{34}) 
+ (1 - P_{34}) U\; G_0 \; t \;G_0 \; {\cal{U}}^{11} ,
\end{align}
\end{subequations} 

\noindent where ${\cal{U}}^{11}\; ({\cal{U}}^{21})$ is the transition operator for
$1\!+\!3\!\to \!1\!+\!3 \; (1\!+3\!\to \!2\!+\!2)$, $t$ the $NN$ t-matrix and
$G_0$ the four free particle Green's function. The operators $U$ and $\tilde U$ are
respectively $1+(3)$ and $(2)+(2)$ subsystem transition operators given by 
\begin{subequations} \label{eq:UetildeU}
\begin{align} \label{eq:U}
U = P \,G_0^{-1} + P\, t\, G_0 \, U ,
\\
\label{eq:tildeU}
\tilde U = \tilde P \,G_0^{-1} + \tilde P\, t\, G_0 \, \tilde U, 
\end{align}
\end{subequations}

\noindent where the permutation
operators $P$ and $\tilde P$ are given by
$P = P_{12} P_{23} + P_{13} P_{23}$, and  $\tilde{P} = P_{13} P_{24}$.
Defining $|\phi_1\rangle$ and $|\phi_2\rangle$ the initial/final $(1+3)$ and $(2+2)$
states that satisfy the equations
\begin{subequations} \label{eq:phi1e2}
\begin{align} \label{eq:phi1}
|\phi_1 \rangle = G_0\, t\, P\,|\phi_1 \rangle,
\\
\label{eq:tildephi2}
|\phi_2 \rangle = G_0\, t\, \tilde P\,|\phi_2 \rangle, 
\end{align}
\end{subequations}
the matrix elements 
$\langle \phi_\alpha|T^{\alpha\beta}|\phi_\beta \rangle = S_\alpha
\langle \phi_\alpha|{\cal{U}}^{\alpha\beta}|\phi_\beta \rangle $ 
with $S_1 = 3$ and $S_2 = 2$ lead to the transition amplitudes
from which one calculates observables. After partial wave decomposition we
solve a three-variable integral equation with no approximations beyond the
usual discretization of continuous variables in a finite mesh.    

\section{RESULTS}

In order to calibrate the accuracy of our new numerical algorithm we calculate first
the negative eigenvalues of the kernel that correspond to the $^4{\rm He}$ binding
energy. The Coulomb interaction is included through screening and charge dependence
in the $NN$ interaction taken in isospin $T = 0$ alone. In Table 1
we show the convergence of the results for the binding energy in terms of
two-nucleon total angular momentum $I$. The calculation with
$I \leq 6$ includes  1057 channels and clearly shows the accuracy of the numerical
method we use. Results of ref.~\cite{Vivnucl-th} for CD Bonn include total
isospin $T=1$ and $T=2$ states,  whereas they are neglected in our calculations.

\begin{figure}[h] 
\begin{center}
\includegraphics[scale=0.65]{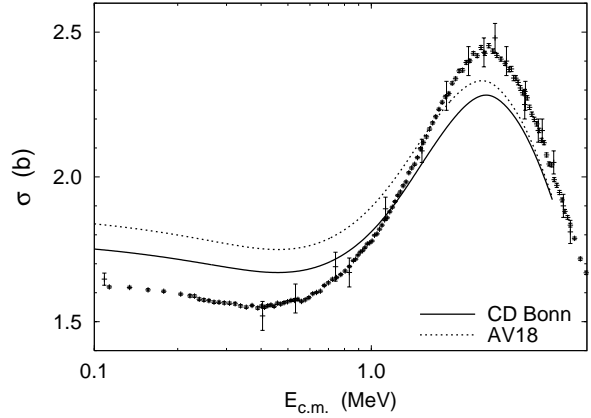}
\end{center} 
\caption{Total cross section for $n\Hh$ scattering versus center of mass energy
for different realistic interactions.}
\label{fig:nttot.eps}
\end{figure}

\begin{table} [h] \label{tab:1}
\centering 
\caption{$\alpha$-particle binding energy as function
of the total two-nucleon angular momentum $I$.
Predictions for AV8' do not include Coulomb. } 
\begin{tabular}{l|*{6}{c}|c} 
& $I \le 1$ & $I \le 2$ & $I \le 3$ & $I \le 4$ & $I \le 5$ & $I \le 6$ 
&  refs.~\cite{Nog01,Kam01,Lazaus04,Vivnucl-th} \\ \hline
AV8' & 23.08 & 25.16 & 25.69 & 25.85 & 25.90 & 25.91  
&  25.90 - 25.94 \\
AV18 & 22.30 & 23.75 & 24.15 & 24.20 & 24.23 & 24.24  &  24.22 - 24.25   \\
CD Bonn & 25.03 & 25.95 & 26.07 & 26.10 & 26.11 & 26.11 
&  26.13 - 26.16  \\
INOY04 & 28.68 & 29.09 & 29.10 & 29.11 & 29.11 & 29.11  &  29.11   \\
\end{tabular}
\end{table}
\begin{table}[h] \label{tab:2}
\centering
\caption{$n\Hh$ phase shifts and mixing
parameter at $E_n$ = 3.5
MeV together with total cross section.}
\begin{tabular}{lccccccccc} \hline
 & 0$^+$  & 0$^-$ & 1$^+$  & \multicolumn{3}{|c|}{1$^-$} &  2$^-$ &
$\sigma_t(b)$ &
\\
\hline
 & ($^1$S$_0$) & ($^3$P$_0$) & ($^3$S$_1$)  & ($^3$P$_1$) & ($^1$P$_1$) & $\delta$
&($^3$P$_2$) & &
\\ 
\hline \cr
AV 18 & -66.07 & 20.72 & -58.44 & 40.08 & 20.71
& -44.50 & 42.48 & 2.329 & this work \\ 
& -66.5\,\, & 20.9 \,\,&
-58.5 \,\,& 37.3 \,\,&
20.7\,\,& -43.5\,\,&
41.0 \,\,& 2.24\,\, & ref. [6]\\
& -66.3\,\, & 20.6 \,\,&
-58.7 \,\,& 38.6 \,\,&
20.5\,\,& -45.5\,\,& 
40.1 \,\,& 2.24\,\, & ref. [6]\\ 
& -63.7\,\, & 27.5 \,\,&
-58.3 \,\,& 44.8 \,\,&
24.7\,\,&  &
44.2 \,\,& 2.51\,\, & ref. [6]\\ [0.9ex]              
CD-Bonn & -64.63 & 18.97 & -57.41 & 39.44 & 20.19
& -44.95 & 42.45 & 2.282 & this work \\ [0.5ex]
 \hline
\end{tabular}
\end{table} 
Next we show results for $n\Hh$ scattering by solving Eqs.~(\ref{eq:U11}) and
(\ref{eq:U21}) in isospin $T = 1$. The phase shifts at $E_n = 3.5$ MeV are shown in
Table~\ref{tab:2} together with the total cross section $\sigma_t$. The present work
confirms previous calculations by the Grenoble and Pisa groups (second and third lines
respectively) compiled in ref.~\cite{Lazaus05} together with the results by one of us
(fourth line) where the AV18 interaction was represented in rank one. The small
differences between the phases in the first three lines are probably due to higher 
$NN$ partial waves $(I \leq 4)$ that are included in the present calculation
together with $\ell_y, \ell_z \leq 4$. In spite of the improved accuracy, both
AV18 and CD Bonn interactions lead to cross sections in the resonance region
that fail to reproduce the experimental data. This is shown in
Fig.~\ref{fig:nttot.eps}.
  
Finally we use the method developed for $pd$ elastic scattering
\cite{Delt05,Delt05a}  to include the Coulomb interaction between the protons.
This means using the  two-potential formula
\begin{equation} \label{eq:Phi1}
\langle \phi_{_1}  \vec p_f|T^{11}|\phi_{_1}  \vec p_i \rangle =
\langle \vec p_f |t_C| \vec p_i  \rangle  
 + \lim_{R \to \infty} \left\{ Z_R^{- \frac{1}{2}} (p_f)  \langle \phi_{_1} 
\vec p_f| \left[ T^{11}_{(R)} - t_R \right] |\phi_{_1}  \vec p_i 
\rangle   Z_R^{-\frac{1}{2}} (p_i)  \right\},
\end{equation}
where the first term is the long range Coulomb amplitude between the proton and
the center of mass of $^3{\rm He}$ and the second term represents the Coulomb
modified nuclear short range amplitude that results from the difference between
the matrix elements of the AGS operator $T^{11}_{(R)}$ calculated with
screened Coulomb  between the three protons and the two body $p^3{\rm He}$
screened Coulomb t-matrix $t_R$ after renormalization with
$Z_R^{-\frac{1}{2}}$.  
 
\begin{figure}[h]
\begin{center}
\includegraphics[scale=0.6]{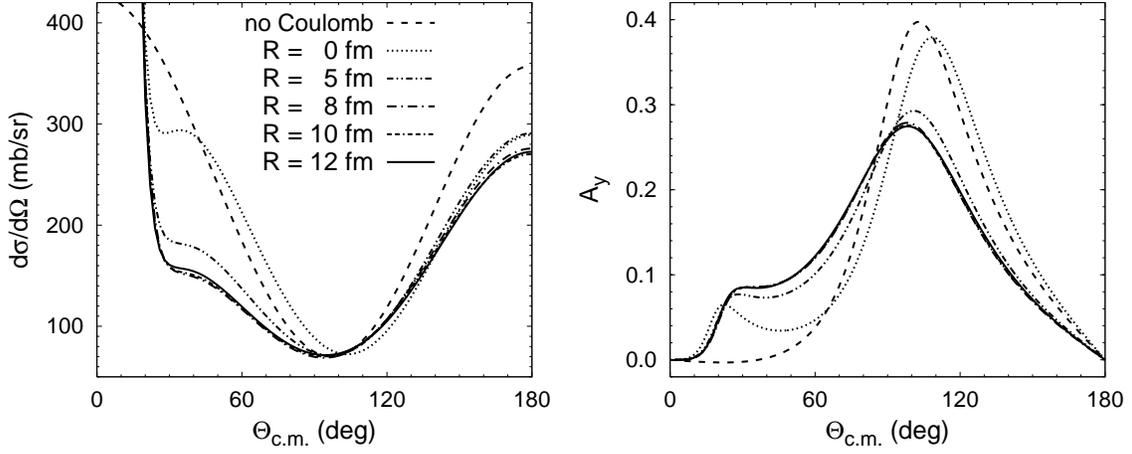}
\end{center} 
\caption{Convergence of $p\He$ observables at $E_p = 4$ MeV with the screening
radius $R$ of the Coulomb potential.}
\label{fig:Rp3He4.eps}
\end{figure} 

\begin{figure}[!]
\begin{center}
\includegraphics[scale=0.6]{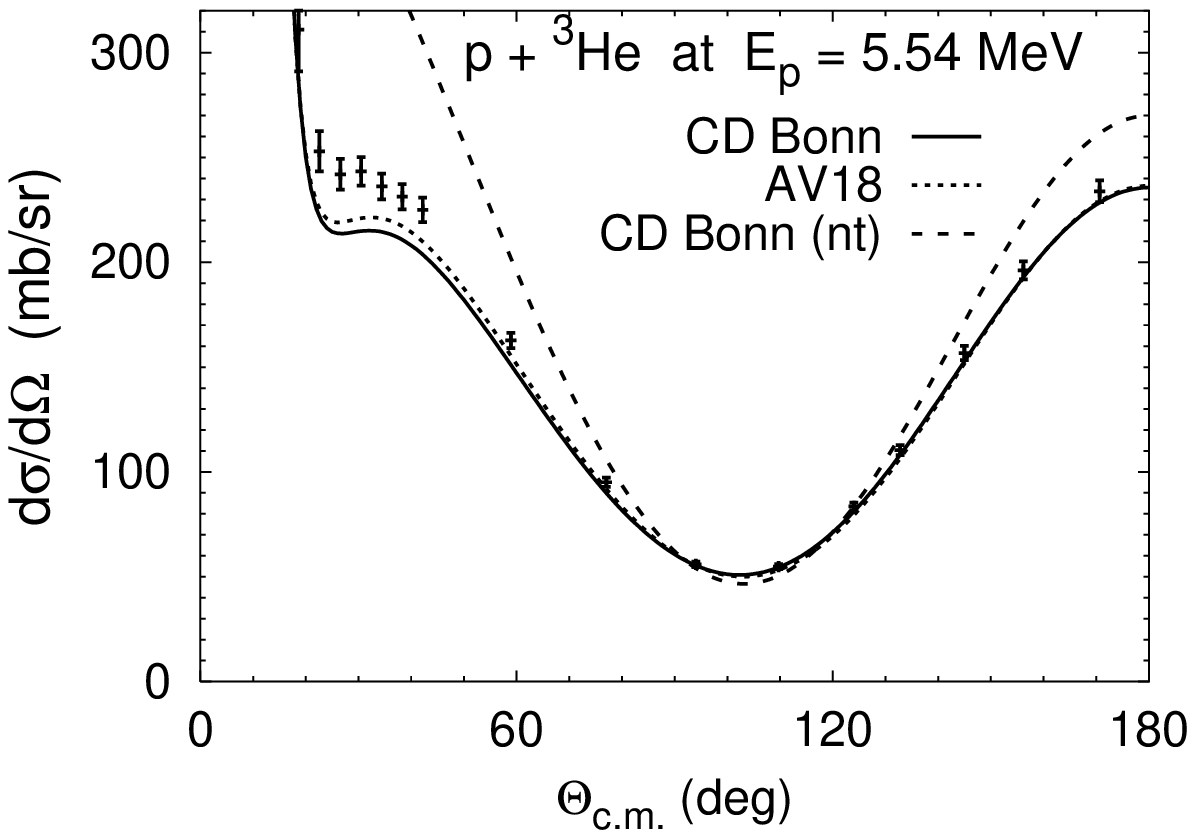}
\includegraphics[scale=0.6]{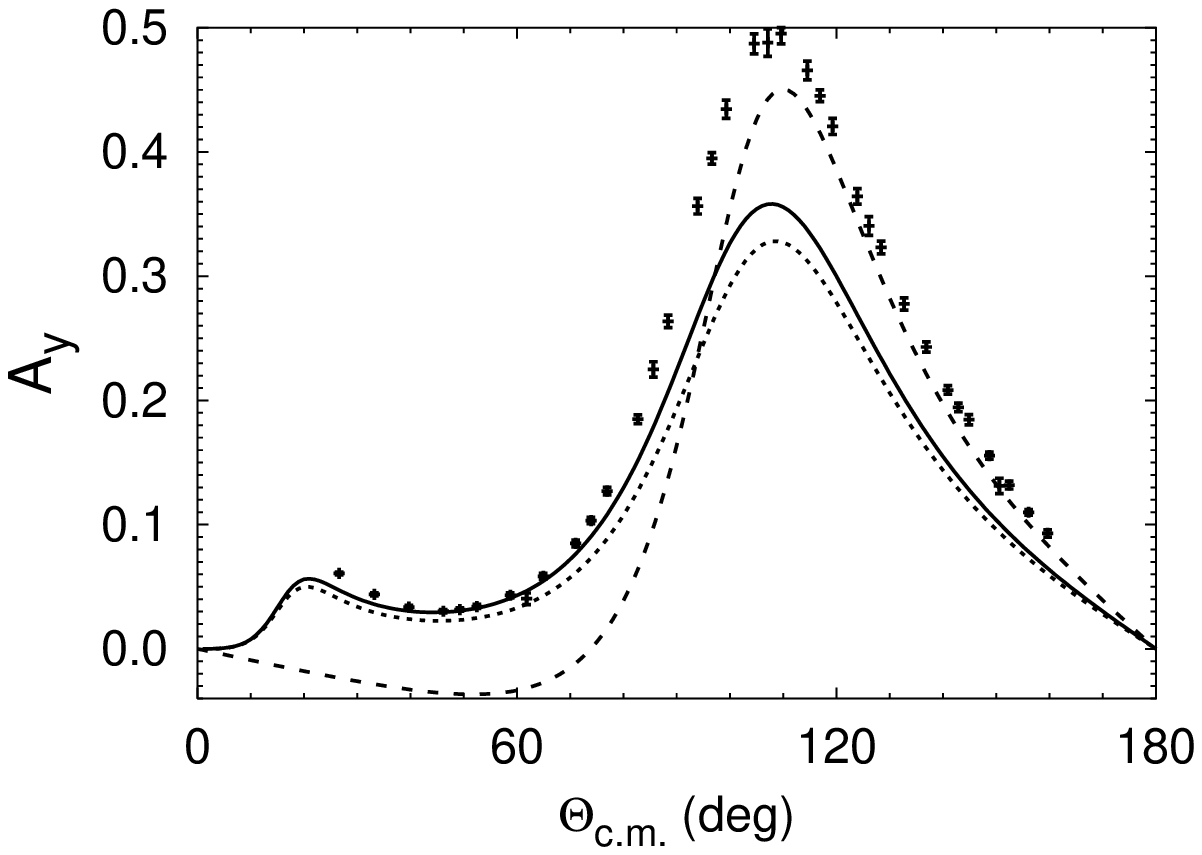}
\end{center} 
\caption{Differential cross section and nucleon analyzing power
for $p\He$ scattering at $E_p = 5.54$ MeV
using CD Bonn (solid) and AV18 (dotted) potentials plus the Coulomb 
interaction between all three protons. The dashed line corresponds to 
CD Bonn alone. Experimental data are from refs.~\cite{mcdonald:64,alley:93}.}
\label{fig:ph5.5a}
\end{figure}

\begin{figure}[!]
\begin{center}
\includegraphics[scale=0.6]{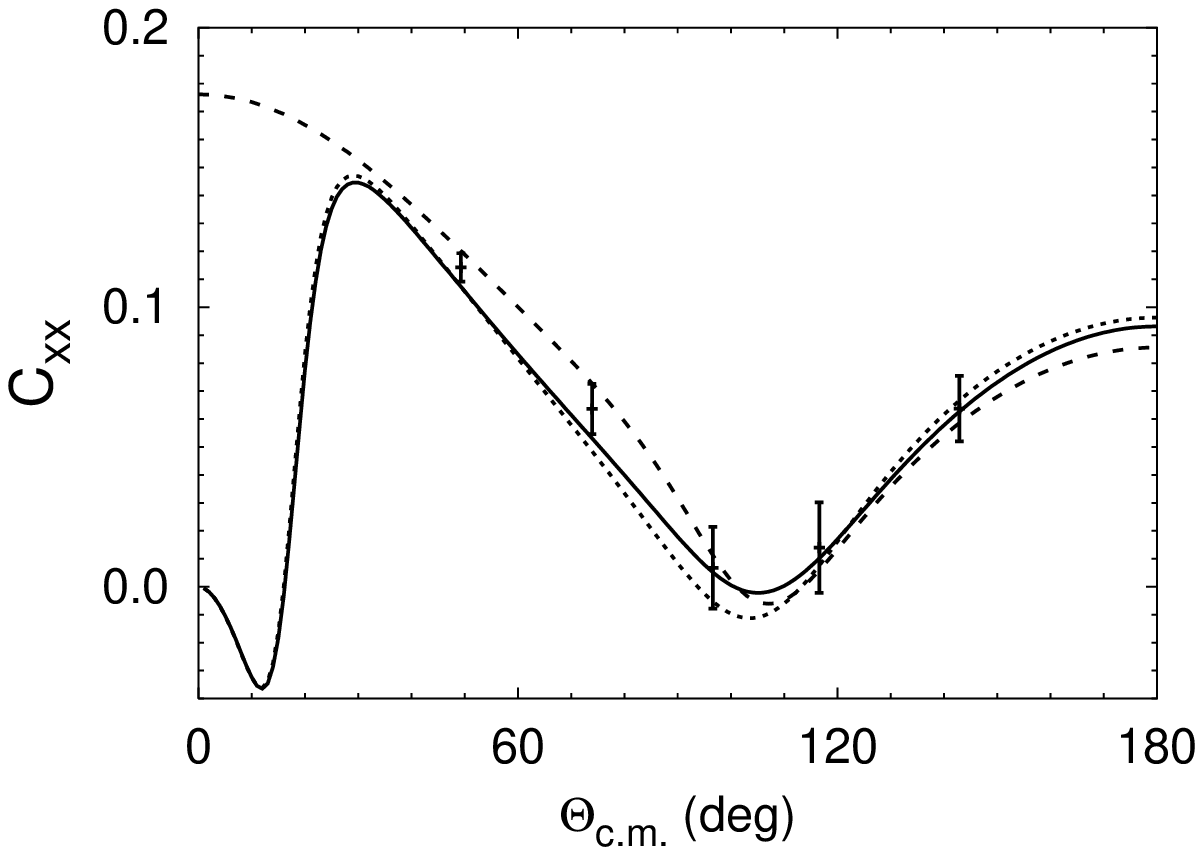}
\includegraphics[scale=0.6]{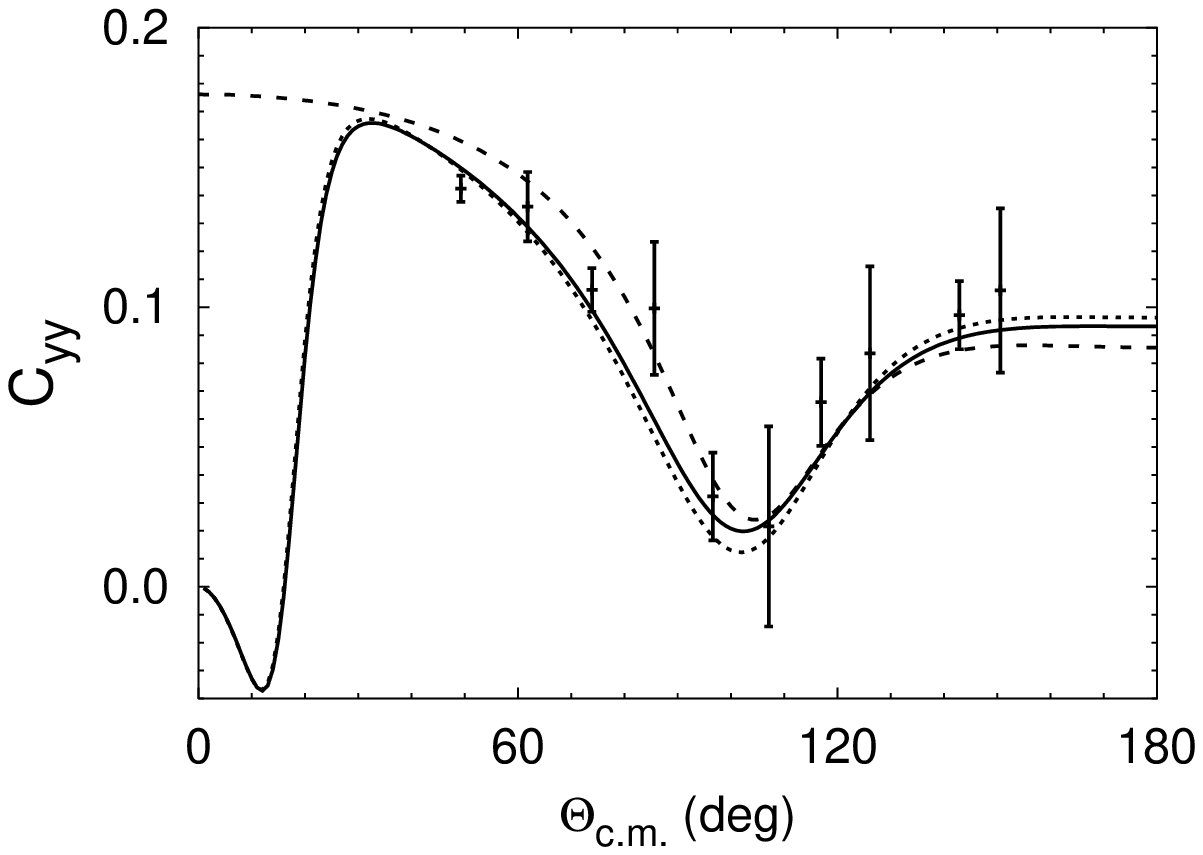}
\end{center}
\caption{Spin correlation coefficients  
for $p\He$ scattering at $E_p = 5.54$ MeV.
Curves as in Fig.~\ref{fig:ph5.5a}.
Experimental data are from ref.~\cite{alley:93}.}
\label{fig:ph5.5b}
\end{figure}

The convergence of the calculation with the screening radius $R$ is shown in
Fig.~\ref{fig:Rp3He4.eps}. For $R = 12$ fm we get converged results for both
observables. Further results are shown in Figs.~\ref{fig:ph5.5a} and  \ref{fig:ph5.5b} 
for the $d\sigma/d\Omega$, $A_y$, $C_{xx}$ and $C_{yy}$ at $E_p = 5.54$ MeV. 
The calculations
show that there are large Coulomb effects, and greater sensitivity to the $NN$
interaction compared to what is observed in $pd$ scattering at low energy.
The calculated analyzing power $A_y$ shows a large discrepancy with the
data but $C_{xx}$ and $C_{yy}$ are in reasonable agreement with experiment.

\hspace{\fill}
%


\end{document}